\documentclass[10pt,journal,compsoc]{IEEEtran}               

\ifCLASSOPTIONcompsoc
  \usepackage[nocompress]{cite}
\else
  \usepackage{cite}
\fi

\ifCLASSINFOpdf
   \usepackage[pdftex]{graphicx}
\else
 \fi

\graphicspath{{figures/}{pictures/}{images/}{./}} 

\usepackage{microtype}                 
\PassOptionsToPackage{warn}{textcomp}  
\usepackage{textcomp}                  
\usepackage{mathptmx}                  
\usepackage{times}                     
\usepackage{cite}                      
\usepackage{tabu}                      
\usepackage{booktabs}                  

\usepackage{xcolor}

\usepackage[ruled,vlined]{algorithm2e} 
\usepackage{amsmath}
\usepackage{amssymb,mathtools}
\usepackage{enumerate}
\usepackage[english]{babel}
\usepackage[utf8]{inputenc}
\usepackage{amsmath}
\newcommand\argmin{\mathop{\operatorname{arg\,min}}\limits}

\usepackage{gensymb}
\usepackage{multirow}
\usepackage[noend]{algpseudocode}

\begin{document}

\title{Geometry-Aware Planar Embedding\\ of Treelike Structures}

\author{Ping~Hu, Saeed~Boorboor, Joseph~Marino, and Arie~E.~Kaufman, \IEEEmembership{Fellow,~IEEE}
\IEEEcompsocitemizethanks{\IEEEcompsocthanksitem Hu, Boorboor, Marino, and Kaufman are with the Department of \protect\\
Computer Science, Stony Brook University.\protect\\ 
E-mail:~\textit{\{pihu, sboorboor, jmarino, ari\}}@cs.stonybrook.edu}

\thanks{Manuscript received xx xxx. 202x; accepted xx xxx. 202x. Date of Publication xx xxx. 202x;}}



\IEEEtitleabstractindextext{%
\begin{abstract}
The growing complexity of spatial and structural information in 3D data makes data inspection and visualization a challenging task. We describe a method to create a planar embedding of 3D treelike structures using their skeleton representations. Our method maintains the original geometry, without overlaps, to the best extent possible, allowing exploration of the topology within a single view. We present a novel camera view generation method which maximizes the visible geometric attributes (segment shape and relative placement between segments). Camera views are created for individual segments and are used to determine local bending angles at each node by projecting them to 2D. The final embedding is generated by minimizing an energy function (the weights of which are user adjustable) based on branch length and the 2D angles, while avoiding intersections. The user can also interactively modify segment placement within the 2D embedding, and the overall embedding will update accordingly. A global to local interactive exploration is provided using hierarchical camera views that are created for subtrees within the structure. We evaluate our method both qualitatively and quantitatively and demonstrate our results by constructing planar visualizations of line data (traced neurons) and volume data (CT vascular and bronchial data).
\end{abstract}

}

\maketitle

\IEEEdisplaynontitleabstractindextext
\IEEEpeerreviewmaketitle
\IEEEraisesectionheading{\section{Introduction}\label{sec:introduction}}
\IEEEPARstart{M}{any} critical biological systems, such as the vascular, respiratory, and nervous systems, consist of structures that express a treelike topology. 
Recent advancements in image acquisition technology coupled with techniques for 3D reconstruction and visualization~\cite{preim2016survey} have enabled scientists to examine the attributes of these intricate structures at a finer resolution, such as their geometry, branching pattern, curvature, and flow capacity. 
However, 3D exploration becomes challenging as the branching morphology of an object grows in complexity. 
This is primarily due to data density clutter, visual occlusions, and at times, the inability to seamlessly manipulate camera orientations within the visualization application during inspection.
Due to these challenges, domain scientists often prefer a simplistic schematic representation~\cite{kreiser2018survey,won2009uncluttered}. 
This has motivated us to develop a method for generating planar embeddings of treelike topologies, such that there are no visual occlusions and the 2D layout preserves the geometric attributes of the original structure.

Given the significance of treelike structures in life sciences, planar visualizations can be very beneficial in clinical settings.
For instance, it can be incorporated in medical visualization tools as a guidance map to facilitate 3D inspection. 
By presenting a single-view representation, physicians can track their observation coverage as well as get a sense of the current 3D view location with respect to the entire structure.
Moreover, akin to the common practice of hand-drawing schematic representations, a planar illustration capturing anatomical and physiological attributes can be used for medical record keeping. 
Using these images, experts can effectively annotate and communicate key features to other collaborators or use it to juxtapose the data with other records to track and detect variations. 
However, to ensure the effectiveness of these representations, it is important that the planar embedding preserves the geometry of the original structure and, specifically for treelike objects, avoid misleading overlaps and branching distortions.

In this work, we present a method that utilizes the 3D skeleton of a treelike object to construct a planar embedding of its topology. 
To maintain geometric context, our method progressively divides a tree structure into smaller segments of single branches and computes local projection planes that best preserve the topology of the points in each segment as well as the topology of points that connect adjacent segments, when mapped onto a 2D view. 
Our approach is in contrast with existing geometry-coherent planar visualization methods~\cite{marino2015planar,won2012uncluttered} that resolve embeddings based on a single, user-defined global projection plane.
To this end, we introduce a novel viewpoint determining algorithm, based on particle swarm optimization~(PSO)~\cite{Kennedy:1995:PSO,Jamian:2014:GPS}, that automatically computes the best projection plane for a set of 3D edges (lines connecting two 3D points).
PSO is a computational technique that solves an objective function by distributing a swarm of particles in space and determines a candidate solution based on a given measure of quality. 
For our algorithm, we formulate a PSO that minimizes loss in geometry information -- specifically length, curvature, and branching angle -- when projected onto a 2D plane.
Using 2D projections of skeleton branches that are reflective of the original 3D morphology, we compute the set of bending angles that will subsequently be used to generate a planar embedding of the treelike structure.
Similar to the viewpoint algorithm, we introduce a GPU accelerated energy function, based on PSO, that generates an embedding with no intersections and with minimal variations from the original 3D edge lengths and bending angles.
Realizing that dimension reduction cannot possibly preserve all attributes, in our implementation we allow users to prioritize the preservation of length and angle attributes by adjusting their weights in the planar embedding energy function.
Our planar embedding algorithm pipeline is shown in Fig.~\ref{fig:pipeline}.

Moreover, we extend our viewpoint algorithm to facilitate 3D exploration of treelike objects. 
Essentially, visualizing 3D objects on a computer screen is analogous to 2D projection and it can be a cumbersome task to manually find a view that avoids visual occlusions.
Particularly, for an object with open endings, such as a treelike structure, its shape and morphology are not only perceived by observing its continuous connected edges but also by the spatial relationship between indirectly connected segments.
Our viewpoint selecting method handles this characteristic better than the existing methods for mesh-based objects. 
Finally, for an effective 3D exploration, our technique automatically generates a smooth camera path using a series of B\'ezier curves.

We summarize the contributions of our work are as follows:
\begin{itemize}
\item A planar visualization method with interactive weight adjustment for treelike data to achieve an occlusion-free planar view while preserving the local geometry information (that is, the node angles and edge lengths) of each segment of the treelike structures, without requiring any initial viewpoint by the user. 
    \item A planar embedding using automatically generated local viewpoints around the structure to preserve both the overall global shape and the local shape of individual branches.
    \item An interactive component where the user can manually reposition skeleton nodes and our method will update the final embedding.  
    \item A hierarchical navigation of the original 3D structure based on our principal viewpoint collection.
\end{itemize}



\section{Related Work}
\label{sec:related}
The related works is divided in two areas: (1) planar visualization techniques and their applications, and (2) camera view generation.

\subsection{Planar Visualization Techniques}
\label{Sec:related_works_planar}
Formal user studies have shown that using 2D data representations can be more effective in carrying out tasks that require spatial identification and precision, than 3D visualization~\cite{tory2007spatialization, laidlaw2005comparing}.
This evaluation is particularly important for clinical applications that involve inspection and diagnosis of medical data, and therefore, there has been a great effort in introducing methods for constructing their planar visualizations.
In this section, we discuss techniques for the planar visualization of tubular objects expressing a treelike topology. For a complete review on varying biomedical data, we refer the reader to a survey by Kreiser et al.~\cite{kreiser2018survey}. 

Broadly, planar embedding of tubular structures can be categorized into techniques that visualize the surface of the objects (interior or exterior), or its internal volume.
Most often, surface flattening methods present a single non-occluded view by mapping the surface to a 2D plane in a linear arrangement~\cite{bartroli2001nonlinear, hong2006conformal,liang2009extracting,Eulzer:2021:ACF}.
To visualize the surface of treelike structures, Lichtenberg and Lawonn \cite{Lichtenberg:2020:PFE} have presented a texture coordinate generation technique that can be applied to vascular contour parameterization, feature extraction, and structure segmentation.
For a branching topology, Zhu et al.~\cite{zhu2005flattening} have proposed partitioning the vessel anatomy into segments based on bifurcation points and subsequently map the straightened surface segments onto a plane.
Moreover, Eulzer et al.~\cite{Eulzer:2021:ACF} have developed a cutting and flattening method for vascular geometry, mapping the 3D surface onto 2D domain as a single patch.
Though these methods preserve local geometric properties, the overall context of the object's shape is lost.

To overcome this limitation, Won et al.~\cite{won2012uncluttered} have developed a method that generates a schematic representation of branching structures by defining an optimization problem using simulated annealing.
The object is first divided into bounding boxes, and an optimization function then solves for a spatial configuration of the boxes while avoiding overlaps and keeping its topology similar to an input projection view. 
Marino and Kaufman~\cite{marino2015planar} have described an embedding approach that determines bending angles of a treelike structure by projecting its skeleton onto an input viewing plane.
Starting from a modified radial layout, angular positions of the skeletal nodes are iteratively adjusted until the bending angles are recovered to the best possible extent, while avoiding intersections.
We consider these two methods to be closest to our goal. 
While these approaches depend on an input global viewing direction, which may not truly represent the local morphology, our approach does not require a manual viewpoint selection to represent the global shape.
In contrast, our method aims to preserve local morphology by projecting segments to automatically computed local planes. Particularly compared to~\cite{marino2015planar}, which uses a single viewpoint, we create a viewpoint for each branch so that both global and local shape are preserved to the greatest extent possible. 

For visualizing the internal volume of vessels, curved planar reformation (CPR) is a commonly used technique and has been developed to map the lumen of a blood vessel on a 2D plane~\cite{kanitsar2002cpr}.
To solve for a tree topology, later works suggest applying CPR to individual branches and then compositing the results to generate a final visualization~\cite{kanitsar2003advanced, kanitsar2006diagnostic}, albeit being susceptible to occlusions.
For removing occlusions, constraints on spatial relations have to be relaxed by stretching or straightening the vessels, at the cost of distorting geometric context.
Many extensions to CPR have been presented that improve rendering of the interior volume and enhance visibility in the presence of other vessels or objects~\cite{lee2006tangential,auzinger2013vessel,mistelbauer2012centerline,mistelbauer2013vessel}.
Since CPRs also embed the volume surrounding the vessels, the different techniques trade off between overlapping vessels or losing isometry.
Our work only embeds the segmented object and hence does not consider preserving the geometric context of the surrounding structures.

Particularly in biomedicine, tree structures often have a direct spatial correspondence and hence, to facilitate comprehension, developing applications for its visualization has been an ongoing area of research~\cite{hahn2001visualization}.
VesselMap~\cite{tao2016vesselmap} has been developed as an interactive visual analysis framework that presents a view‐dependent 2D graph layout optimized for visualizing cerebral aneurysm and parent vasculature.
NeuroLines~\cite{al2014neurolines} provides a visual interface to help neuroscientists analyze the connection between individual trees to identify appropriate attributes for neurite comparison.
HemoVis~\cite{borkin2011evaluation} has been developed as an interactive visualization application for heart disease diagnosis that reconstructs a 2D tree diagram representation of coronary artery trees.
While effective for comparison tasks, these diagrammatic representations are foreign to physicians, who are accustomed to the original shapes of the objects~\cite{angelelli2011straightening}.
To encode a correspondence between the 3D structure and its 2D representation, Lichtenberg et al.~\cite{Lichtenberg:2019:DFV} have proposed using a colored binary tree.  
Moreover, their work also visualizes the distance field on the both views, based on the requirements of different application scenarios.
We demonstrate that among other representations, our embedding result can be used as a visual interface for inspecting the topology of complex treelike objects.

\subsection{Camera Viewpoint Selection}
\label{sec:relatedWork_camera}
Given a scene content, a view is determined by the configuration of the camera.
Through a good viewpoint, fundamental data can be perceived and significant objects can be recognized.
Since viewpoint selection is a broad research topic, we focus on the relevant works adopting low level attributes of the view content and the viewpoint selection metrics.
Bonaventura et al.~\cite{Bonaventura:2018:SVS} have summarized the pioneering viewpoint selection methods for polygonal data.
Kamada and Kawai have proposed a basic rule for finding a good view, maximizing the projected lines on each object~\cite{Kamada:1988:SMC}.
Inspired by information entropy theory~\cite{feixas1999information,I2002information}, many works have designed  methods that formulate the problem in entropy space. Arbel et al.~\cite{Arbel:1999:VSN} measure object recognition in a monochrome view, which studied the view selection problem based on Shannon entropy maps.
V\'{a}zquez et al.~\cite{vazquez2001viewpoint} have further modeled view selection as an object visibility maximization problem by transforming view finding into a visibility probability space.
The attributes used are the projected area of each unit in the mesh-based objects.  
Stoev et al.~\cite{Stoev:2002:CSA} have designed a method for automatic camera positioning in time, considering projected area and scene depth.
Podolak et al.~\cite{Podolak:2006:PRS} automatically choose a good viewpoint by minimizing the visible object symmetry.

View selection methods have also been investigated for volume data~\cite{Takahashi:2005:FDA}. 
Bordoloi et al.~\cite{bordoloi2005view} have developed a viewpoint selection method for static volume visualization and their method provides a set of representative views for the volume that partition the view space into several subsets based on view similarity.
For dynamic volumes, Ji et al.~\cite{Ji:2006:DVS} have formulated a utility function that combines voxel opacity, color, and curvature information.
For terrain data visualization, Stove and Straber ~\cite{Stoev:2002:CSA} have proposed considering maximum depth, in addition to projected area.
Focusing on the camera views inside blood vessel models, Meuschke et al.~\cite{Meuschke:2017:AVS} have developed an automatic viewpoint selection technique for exploring simulated aneurysm data. Based on their technique, Apilla et al.~\cite{Apilla:2021:AAA} have presented a camera path generation method to explore the model through optimal viewpoints showing the interesting regions during the cardiac cycle.
Recently, methods for viewpoint generation are being developed that take perceptual factors into consideration~\cite{ono1988dynamic,Secord:2011:PMV,PH:2019:RSS}.
Secord et al.~\cite{Secord:2011:PMV} have proposed a perceptual model that combines attributes, projected model, and silhouette length of the polygonal objects to predict the users' preferences to the viewpoint of a variety of objects. 

Our method shares some similarity with Bordoloi et al.~\cite{bordoloi2005view} as it locates the best views to maximize the visible geometry attributes reflected on the skeletal data.
However, we generate a series of hierarchical views that discover the details in the treelike structure based on subtree levels.


\section{Algorithm}
\label{sec:alg}

\begin{figure*}[t]
\includegraphics[width=\textwidth,keepaspectratio]{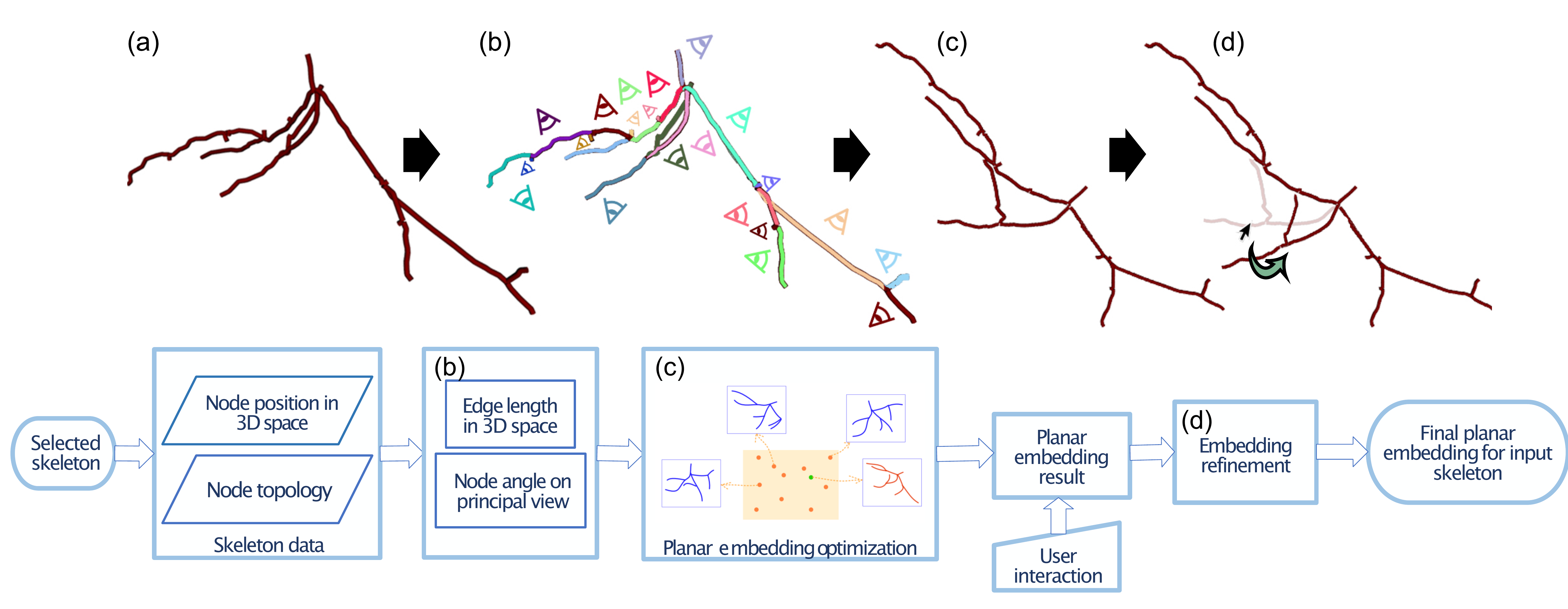}
    \caption{Overview of our planar embedding algorithm.  (a) The original 3D skeleton of a treelike object.  (b) The skeleton is divided into branching segments. Local camera views are determined for each segment which best preserve its local geometric attributes when projected to a 2D plane.  (c) Planar embedding of the 2D skeleton using the projected attributes from the camera views and the radius of each node.  (d) Fine-tuning of the result by the user optionally clicking and dragging nodes. Our algorithm checks again to solve for any intersection introduced by this interaction. }
	\label{fig:pipeline}
 \end{figure*}

Unless visualized in a virtual reality setting, inspecting an object in 3D can be a cumbersome task as it involves manipulating camera orientations to obtain a view with minimal occlusions and for it to be recognizable with respect to the entire structure.
In this work, we address these challenges by introducing two visualization methods. 
First, a technique for planar embedding of treelike objects that preserves the global shape and local branching morphology of the structure, while avoiding intersections.
Second, an automatically generated parallax-aware camera sequence that facilitates an effective 3D navigation of treelike objects.
In elucidating shape preservation, we adopt a description similar to ~\cite{Feragen:2010:GST}, where the overall shape of a treelike object is the combination of each branch segment's local shape, characterized by the set of node angles in the segment, along with the relative placement of the segment's siblings. 
Additionally, we define geometric attributes as branch length, thickness, and branching and bending angle.
Briefly, our embedding algorithm takes as an input a 3D skeleton model of the object and follows three main steps: (1) divide the skeleton into segments and determine local viewpoints for each segment such that the projected points preserves the 3D geometric attributes, (2) generate a planar embedding without intersections, and (3) optionally refine the embedding result based on user interaction.
An overview of these steps is illustrated in Fig.~\ref{fig:pipeline}.

We assume that the skeleton model contains a set of nodes located at points in the structure where there is a change in morphology (such as bends or branching points).
Moreover, each node should have exactly one parent, any number of children, and optionally, a scalar value to represent the radius of the edge (representing line width), that connects the node to its parent.
We define the following terms when referring to the skeleton model:
\begin{itemize}
    \item A \textit{root} node can have several children but has no parent.
    \item A node with multiple children creating a junction in the tree is called a \textit{branching} node.
    \item A node edge length, $l$, refers to the length of the edge connecting the node with its parent.
    \item A sequence of nodes with a single child and the edges connecting these nodes are considered a \textit{segment}.
    \item A \textit{node angle}, $\theta$, is defined as the angle in the clock-wise direction between the edges that connect the node to its parent (node-parent edge) and the parent to the node's grandparent (parent-grandparent edge). 
\end{itemize}

In addition to our embedding technique, we have developed a camera sequence algorithm that generates an optimal 3D navigation path for each subtree of the object, based on its geometry information and spatial relationship with other neighboring branches.
By generating the set of optimal local camera poses, we generate a smooth camera path using B\'ezier curves.

\subsection{Hierarchical View Finding}
\label{sec_hierView}
The goal of finding \textit{good} viewpoints naturally requires defining a set of metrics that can measure the preservation of geometric attributes.
For treelike structures, based on existing literature~\cite{Biasotti:2008:SS,Feragen:2010:GST,Feragen:2011:MST}, we have determined the metrics to be length, curvature, and angle, for geometry information; and radius, branching frequency, and adjacency, for morphology information.
To this end, we present an algorithm to quantitatively optimize these metrics to find good camera viewpoints and subsequently, a method for fast camera path generation using these viewpoints.

\subsubsection{Camera view metric}
Feixas et al.~\cite{Feixas2009Unified} formulate the measure of geometric attributes in a camera viewpoint as:
\begin{equation}
I(v,O) = \sum_{o\in O} {p(o|v) \log \frac{p(o|v)}{p(o)}},
\label{eq:shannon_entropy}
\end{equation}
where $I(v,O)$ represents the geometry attributes carried by an object collection $O$ in a camera view $v$, $p(o)$ is the probability that a corresponding object $o \in O$ is visible and conveys information in a scene, and $p(o|v)$ is the probability of an object $o$ being visible in $v$.
Essentially, this equation computes the sum of attribute information expressed for every object in a certain camera view.

\begin{figure}
    \centering
    \includegraphics[width=\linewidth,keepaspectratio]{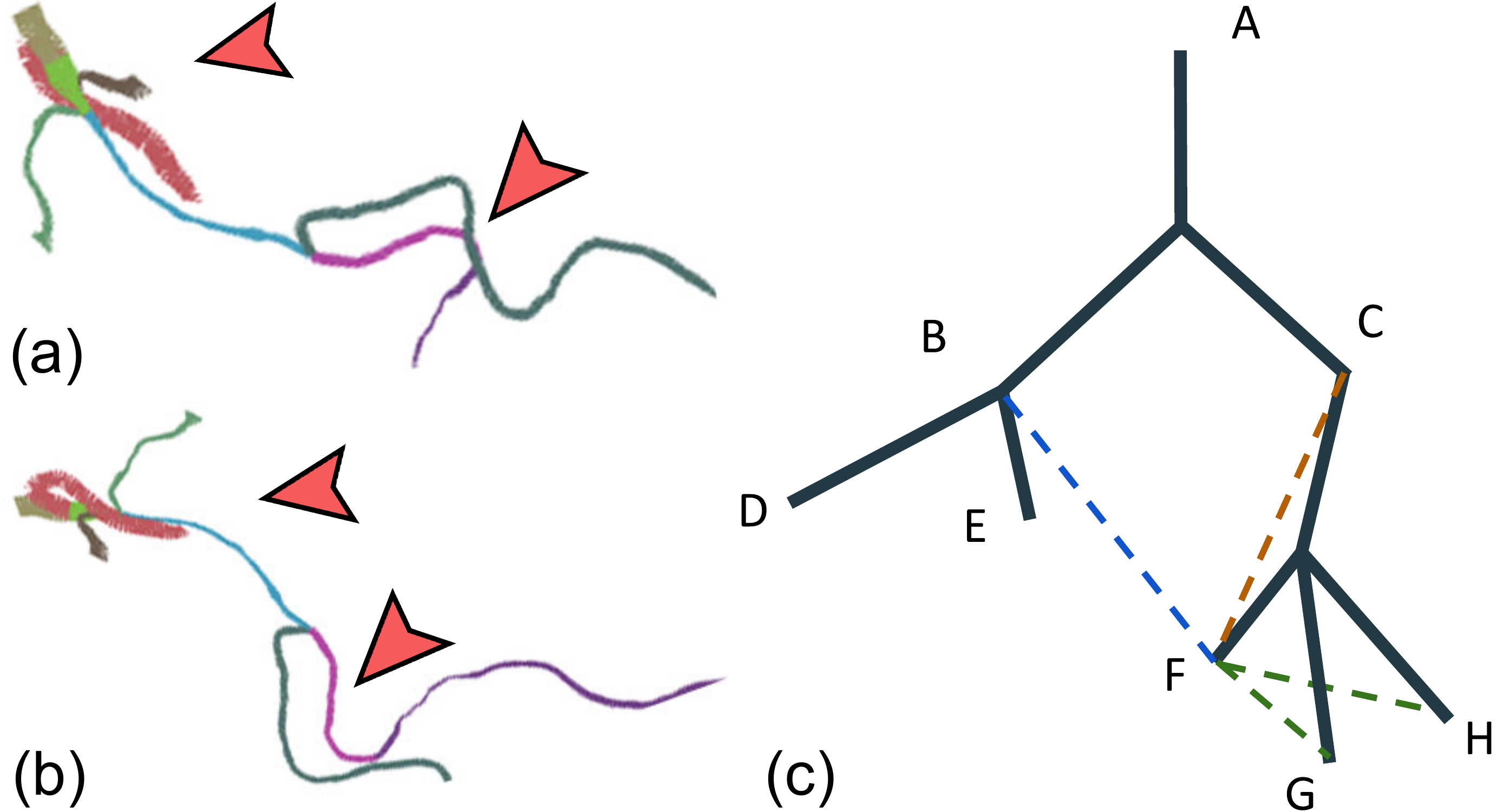}
    \caption{Global best view generated using (a) the method in ~\cite{Takahashi:2005:FDA} and (b) our model. 
    The red arrows show that our \textit{enhanced} tree edges, represented by the dashed lines in (c), result in a more accurate projection of each node's relative placement. The green, red, and blue dashed lines represent three types of imaginary edges for node F.}
    \label{fig:additional_segmentconnect}
\end{figure}

While this model is widely used for viewpoint selection in mesh and volume data (see Sec.~\ref{sec:relatedWork_camera}), it does not always yield a satisfactory camera view for treelike skeleton data. 
This is because treelike skeletons have open endings whose geometry information cannot be fully reflected by preserving only its original length ratio. 
To address this limitation, we have introduced a method that not only considers original 3D edges, but also a set of additional \textit{imaginary} edges as shown in~\ref{fig:additional_segmentconnect}~(c), namely, the connection between each node and (i) its siblings, (ii) its grandparent, and (iii) its parent's siblings.
For simplicity, we refer to the skeleton and the imaginary edges as an \emph{enhanced tree}.

For an enhanced tree, we model the geometry attribute of an edge as the ratio between the edge length $l$ and the total length of all edges.
By extension, given a camera view, the projected attributes coming from an edge is the ratio of its projected length over the sum of projected lengths in the enhanced tree.
The effect of views computed using the additional edges is shown in Fig.~\ref{fig:additional_segmentconnect}~(a) and (b).


\subsubsection{Camera view hierarchy}
\label{sec:viewhierarchy}
We present a camera view hierarchy algorithm that facilitates shape understanding by presenting level-of-detail~(LoD) information using a computed best global view and a series of local view for fine structures.
For each LoD hierarchy, our method determines the best camera position $v_h \in \mathbb{R}^3$.
To this end, we formulate a hierarchical view finding optimization such that the camera viewpoint candidates lie on a spherical surface.
The radius of this spherical search space is proportional to the subtree expansion radius with its center positioned at the weighted center of all subtree nodes, $c^{T}$.
The distance $D$ between the camera and the tree center is defined as $D =\beta R^{T}$, where $R^{T}$ is the radius of the principal view with a weight factor $\beta$.
Based on our experiments, we suggest $\beta$ to be 1.5.
Furthermore, the camera look-at direction is the ray from the camera center to $c^{T}$ and the up direction perpendicular to the look-at direction. 
Thus, to find the optimal camera viewpoint $v_{h}$ for a subtree, $T_{h}$, in the hierarchy level $h$ of the tree, we define the following objective function:
\begin{equation}
        v_h = \argmin_{v_h} E_v \quad \textrm{\textit{,}} \quad E_v = I(v_{h},T_{h})
\end{equation}
The hierarchical viewpoint set, $H$, for a tree structure is denoted as $H=\{\{v_{h}\}|\{v_{h=0}\}\cup\{v_{h=1}\}\cup \cdots \cup\{v_{h=m} \}\}$,
where $1, 2, \cdots, m$ represent the hierarchy levels.
The number of hierarchical viewpoints in each tree level is equal to the subtree number in that level.

\subsection{Skeleton Planar Embedding}
To maintain the 3D geometry of a structure in a single planar view, we first determine a method to best preserve its edge lengths and the node angles.
To achieve this, we design a scheme for projecting the nodes onto a principal plane using our viewpoint selection method described in Sec.~\ref{sec:viewhierarchy}.

\subsubsection{Node angle}
\begin{figure}
    \centering
    \includegraphics[width = \linewidth]{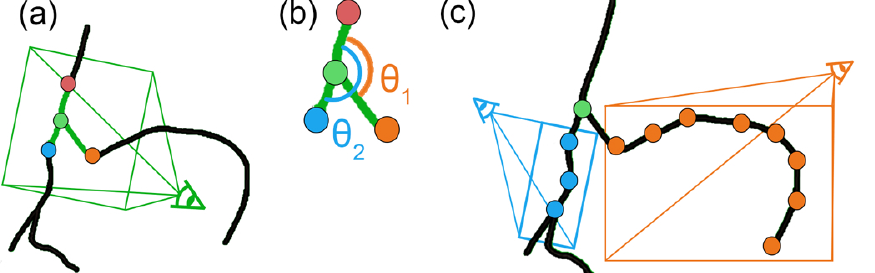}
    \caption{The process of dividing a skeleton tree into segments to compute its planar angles.  (a) A projection plane for a segment consisting of a branching node (green node), its parent (red node), and children (blue and orange nodes) is calculated first.  (b) The branching angle for each child node is the angle between the child node, its parent, and its grandparent.  (c) For each segment, bending angles are calculated by determining a projection plane that preserves the geometric attributes of the entire segment.}
    \label{fig:projection}
\end{figure}
\label{sec:node_angle}

We categorize two sets of node connectivity that determine the target angles for planar embedding: chain and branching.
A chain condition is the set of nodes in a segment where each node has a single child. 
A branching condition is when a node has multiple children.
Given a subtree with a root node that has multiple children, a chain representing a branch of the subtree is collected starting from the root along each node that has a single child, as shown in Fig.~\ref{fig:projection}~(a).
A branching set contains a node, its parent and children, as show in Fig.~\ref{fig:projection}~(b).
To calculate target angles, we first project the relevant set of nodes onto a principal plane and then compute, for each node, the angle formed between its projected node-parent-grandparent edges, in a counterclockwise order.
Given a set of nodes, the principal plane is the plane passing through the subtree center and perpendicular to the look-at direction of the optimal camera view calculated using the method introduced in Sec.~\ref{sec_hierView}.


\subsubsection{Embedding optimization}
\begin{figure}
    \centering
    \includegraphics[width=0.8\linewidth]{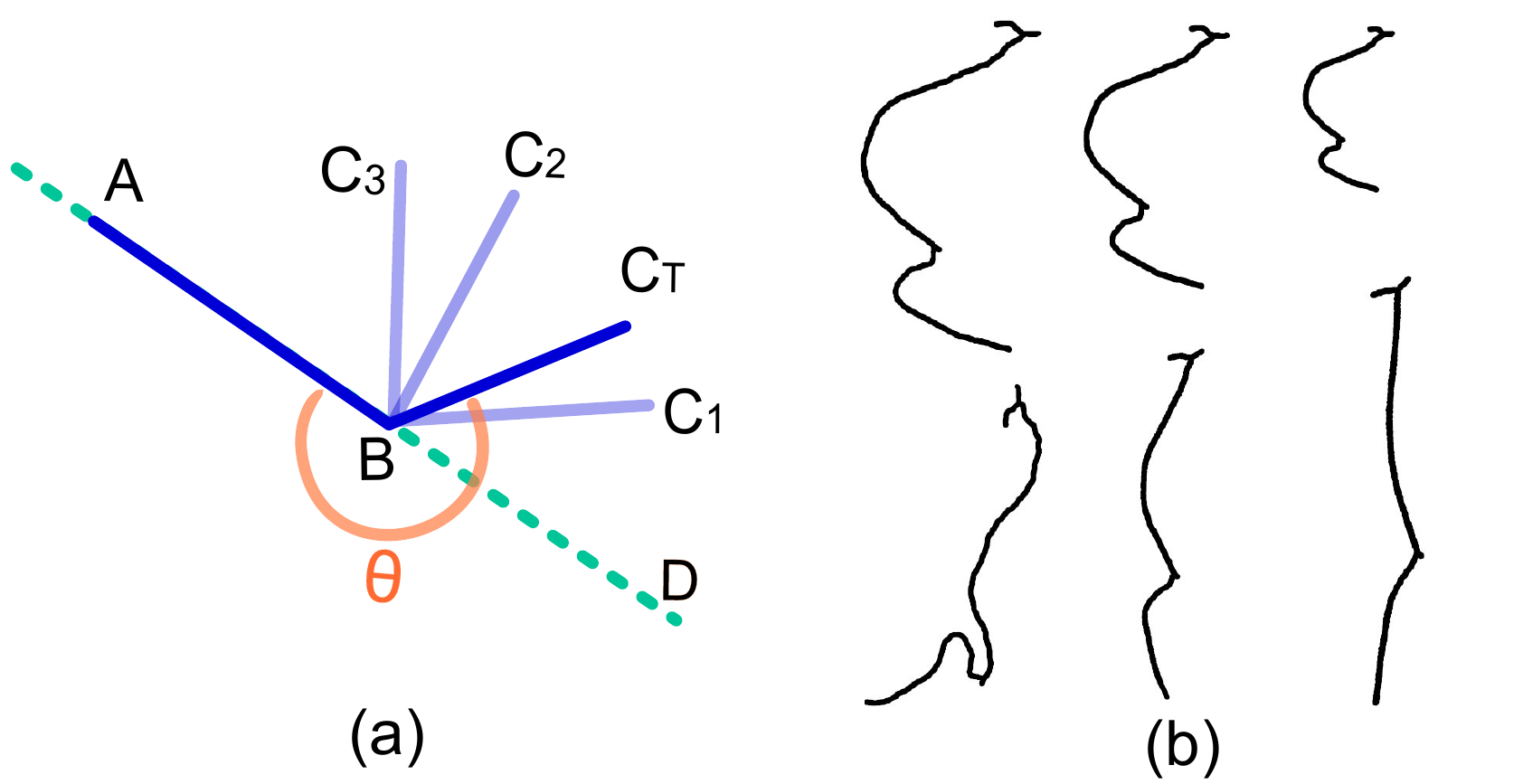}
    \caption{(a) An example of the impact of $r_a$ on $\theta_{e}$ with the node $C$'s target angle $\theta$ ($\angle ABC_T$ in counterclockwise order) as an reflex angle. If $r_a > 0$, node $C$ is positioned at the range between $BD$ and $BC_T$, e.g., $C_1$. If $r_a=0$, node $C$ is at $C_t$. If $r_a < 0$, $C$ is positioned at the range between $BC_T$ and $BA$, e.g., $C_2$ or $C_3$. Thanks to our constrain, $C$ will not be on the $C_T$'s opposite side against the dash line. (b) Segment shapes with different parameters. The top left structure is the original skeleton. The top row shows the changed shape with varying length ratio $r_l$ (left to right: 0.0, 0.2 and 0.5, respectively) and a constant $r_a$ ($r_a=0$). The bottom row shows the result of varying $r_a$ (left to right: -0.02, 0.2 and 0.5, respectively) and $r_l=0$.
} 
    \label{fig:curve_differentparams}
\end{figure}

Using the target angles and edge length, we formulate the planar embedding problem as a UV coordinate optimization procedure. 
In other words, we aim to find a position for each node on a planar view that minimizes loss in target angles and edge lengths.
However, to ascertain geometry preservation of the tree structure, we cannot directly apply a constant minimization factor to all edges and nodes.
For instance, applying a certain shrinking factor to a longer edge can account for a heavier loss in geometry than an edge with a smaller length. 
To address this, we minimize the percentage loss of edge length and node angle, denoted as  $r_{l}$ and $r_{a}$ respectively, in the candidate 2D embedding.
Given $r_l$ and $r_a$, an edge length $l_e = (1+r_l)l$, and a node angle $\theta_{e}$ is
\begin{equation}
    {\theta_{e} =
    \begin{cases}
    \theta + (\pi - \theta)r_a & \text{if $\theta \leq \pi$ and $r_a \geq 0 $ } \\
    (1+r_a)\theta & \text{if $\theta \leq \pi$ and $r_a < 0 $ } \\
    \theta - (\theta - \pi)r_a & \text{if $\theta > \pi$ and $r_a \geq 0 $ } \\
    \theta - (2\pi - \theta)r_a & \text{if $\theta > \pi$ and $r_a < 0 $ }
    \end{cases}
    }
    \label{eq:embedding_angle}
\end{equation}
where $\theta$ is the target node angle calculated in Sec.~\ref{sec:node_angle}.
We constrain $r_l\in(0, 2.0]$ and $r_a \in [-1.0, 1.0]$ in the following optimization step.
Fig.~\ref{fig:curve_differentparams}~(a) shows an example of the $\theta_{e}$ based on different $r_a$ when $\theta \in (\pi, 2\pi)$.
By applying Eq.~\ref{eq:embedding_angle}, the relative bending direction of the nodes will be maintained although the angle value may be changed, as shown in Fig~\ref{fig:curve_differentparams}~(b). 

We optimize a single $r_{l}$ and $r_{a}$ for each segment using:
\begin{equation}
    \begin{split}
        R^{*}_{li},R^{*}_{ai} &= \argmin_{R_{li},R_{ai}} E_p \\
        E_p &= \sum_{i}^{M}{w_l\cdot r_{li}^2 + w_a\cdot r_{ai}^2 + w_X\cdot X(r_{li}, r_{ai})}
     \label{eq:embedding_Energy}
    \end{split}
\end{equation}
where $w_l$ and $w_a$ are the weights for the length and angle loss ratio, respectively, $M$ is the total segment amount, $w_X$ is the penalty weight for intersecting edges, and $X(r_{li}, r_{ai})$ are the corresponding length and angle responsible for the intersection  in the current placement. 
We report our parameter values in Sec.~\ref{sec:eval}.
The resulting node placement of a simple tree structure with different $r_{l}$ and $r_{a}$ settings are shown in Fig.~\ref{fig:curve_differentparams}~(b). 

Since the search space for this optimization is not convex nor differentiable, we apply PSO ~\cite{Kennedy:1995:PSO} to find a local optimal embedding solution. 
PSO is a robust approach that controls parameters without depending on a search space gradient.
In our algorithm, the skeleton placement is determined by the combination of the $\{r_{l}\}$ and $\{r_{a}\}$  ratios, denoted as $R_{l}$ and $R_{a}$ respectively.
Specifically, each swarm particle $P_i$ represents a candidate solution for the placement. 
$R_{l}$ and $R_{a}$ are initialized randomly.
The energy $E_p$ of this embedding is calculated using Eq. \ref{eq:embedding_Energy} that quantifies the goodness of a candidate nodes placement.
A lower energy corresponds to less geometry loss and fewer intersections.
In each iteration, a particle is updated based on its moving velocity $U_i$, current ratio set $\{R_{li}, R_{ai}\}$, historically best ratio set $P_{ig}$, and the ever best particle $P_g$ amongst all particles. This is formulated as: 
\begin{equation}
    P_i^{c+1} = \omega_g \cdot (P_g^{c} - P_i^{c}) + \omega_p \cdot (P_{gi}^{c} - P_i^{c}) + \omega_{inert} \cdot (U_i^{c} + \gamma^{c}) + P_i^{c}
    \label{eq:particle_update}
\end{equation}
where $\omega_g$, $\omega_p$, $\omega_{inert}$ represent the effect of the global best ratio set, the best ratio set within the particle history, and the inertial velocity of the particle itself in the searching space, respectively, $c$ refers to the iteration count, and $\gamma$ is a randomized velocity value to increase the search diversity and accelerate convergence. 
$U_i^c$ is calculated as the difference of the corresponding ratios between the $(c)$th and $(c-1)th$ iteration.
The optimization is terminated when the geometry loss is zero or when the maximal loop limit is reached.
Algorithm \ref{alg:PSO_shaperecovery} provides a pseudocode of our method.

To encourage non-intersecting node placement candidates and to accelerate the optimization convergence, we propose setting $w_X$ to have a value much larger than the other weights.
In evaluating our experiments, we determined this weight to be 1.5 times the sum of the maximal edge length loss and the maximal angle loss. 
Since both the possible maximal angle and edge length loss equals the total number of edges in the tree, a node placement with a single edge crossing will have a higher energy than a placement without any edge crossing, in our implementation.
Moreover, to ensure at least one non-intersecting solution, we additionally generate a radial layout of the tree, using the method in \cite{marino2015planar}, and calculate its $\{R_{l}, R_{a}\}$ as Particle 1's initial values, $\{ Q_l, Q_a \}$.
Thus, in contrast to a general PSO solution, our method guarantees a final non-intersecting placement, no worse than the radial layout.

\begin{algorithm}[ht]
\LinesNumbered
\KwResult \text{Optimal ratio set} $\{R^{*}_{l}, R^{*}_{a}\}$ \text{ denoted as } $P_g$;\\
\KwIn {\text{Randomize} ${R_{l}, R_a, U_i}$ \text{ for each } $P_i$;\\
$P_0 \leftarrow \{r_{lm}=0, r_{am}=0\}, P_1\leftarrow \{Q_{l}, Q_{a}\}$;\\
$P_g \leftarrow P_0, P_{ig} \leftarrow P_i, E_{pi}\leftarrow\infty, c \leftarrow 1, c_{max}\leftarrow 100$}
\While{$E_{p}>0$ \text{ and } $c\leq c_{max}$}{
    \text{For each } $P_i$: \\
    ~~~\text{Generate node placement based on its } $R_l, R_a$ ;\\
    ~~~\text{Update }$E_{pi}$ \text{ using } ${E_p(P_i)}$ \text{ defined in Equation ~\ref{eq:embedding_Energy}};\\
    ~~~$P_{ig}=\min \{ P_{i0}, P_{i1},\cdots, P_{ic}\}$;\\
    End for;\\
    $P_g=\argmin_{P_i}{E_p(P_i)}$;\\
    For each $P_i$:\\
    ~~~ $\{R_l, R_a\}$ \text{ is updated using Equation~\ref{eq:particle_update}};\\
    End for;\\
    $c=c+1$;
 }
\caption{Planar Embedding Using PSO}
\label{alg:PSO_shaperecovery}
\end{algorithm}

\begin{figure*}[ht]
    \centering
    \includegraphics[width=\textwidth,keepaspectratio]{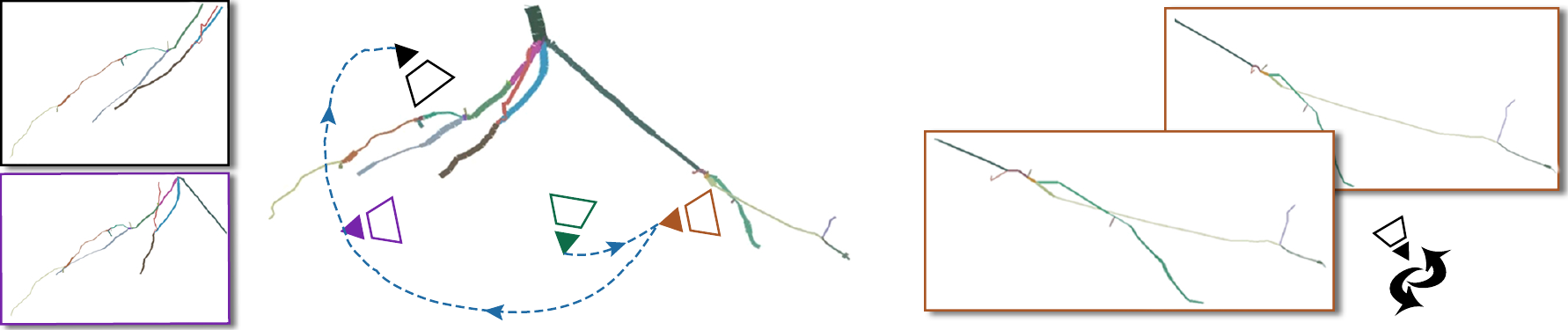}
    \caption{The original structure in 3D space and four screen shots of our automatically generated 3D exploration views. The exploration starts with the best global view and then moves along a smooth B\'ezier path for each subtree's best local view. Screenshots of selected camera views are shown in the insets. We also generate a small camera motion to enhance parallax, as demonstrated using the two insets in the bottom right corner. A video of the exploration is available as supplementary material.}
    \label{fig:exploration}
\end{figure*}

\subsection{Interaction}
\label{sec:interaction}
Researchers have varying preferences with regards to the strictness in geometry attributes preservation.  
For example, a neuroscientist may want to investigate the distance between endpoints of a traced neuron skeleton and as a result, may want to maximize the weight of length preservation. 
Likewise, an anatomist may prefer increasing the weight of angle preservation to observe curvatures along an anatomy. 
To facilitate this, our technique supports the ability to interactively adjust the length and angle weights. 

\begin{figure}
    \centering
    \includegraphics[width=\linewidth,keepaspectratio]{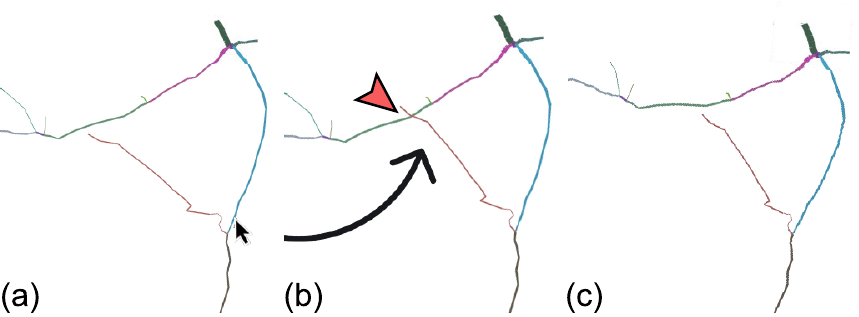}
    \caption{Interactive branch adjustment.  (a) The original embedding result.  (b) The user has repositioned a branch, causing an intersection that is indicated by the red arrow.  (c) The updated embedding result, with the intersection removed.}
    \label{fig:user_interaction}
\end{figure}

Furthermore, our implementation also supports the interactive refining or adjustment of the planar embedding result. 
Specifically, a user can specify any curve or curve segment and rotate it around an anchor node. 
Following the manual adjustment, our method checks to remove any newly introduced intersections.  
An example of this user adjustment is shown in Figure \ref{fig:user_interaction}.
It is worth noting here that the optimization post manual intervention considers the ratios of the user-adjusted segment to be fixed and only recalculates the angle and length ratios of other segments.

\subsection{Exploration in 3D space}
\label{alg:exploration}
It is inevitable to lose some 3D information when reducing the dimensionality from  3D to a planar embedding.
Particularly, information such as the relative position between subtrees and the expansion in 3D space may not be maintained in a 2D view.
To fully uncover the geometry information carried in a 3D structure, our method generates an exploration path based on the hierarchical view set $H$ in Sec.~\ref{sec_hierView}.
To support the effective investigation of geometry information inherent in a 3D structure, we introduce a technique to generate an exploration path based on the hierarchical view set $H$, discussed in Sec.~\ref{sec_hierView}.

This exploration path is composed of multiple groups of camera motion sequences.
Each group includes two motion sequences: a transition from between consecutive principal views and a camera dolly motion to stimulate parallax. 
The camera positions in the transition clip is generated using a quadratic B\'ezier curve connecting one principal view to the next  
and the camera orientation is interpolated using a smooth transition between the two views.
For the dolly motion, the camera is always looking at the center of the subtree and moving along an arc of $90\degree$.
Fig.~\ref{fig:exploration} demonstrates an example of the camera path.
Moreover, please refer to our supplementary video for further examples.

\section{Results}
\label{sec:results}
\begin{figure*}[t]
\includegraphics[width=\textwidth,keepaspectratio]{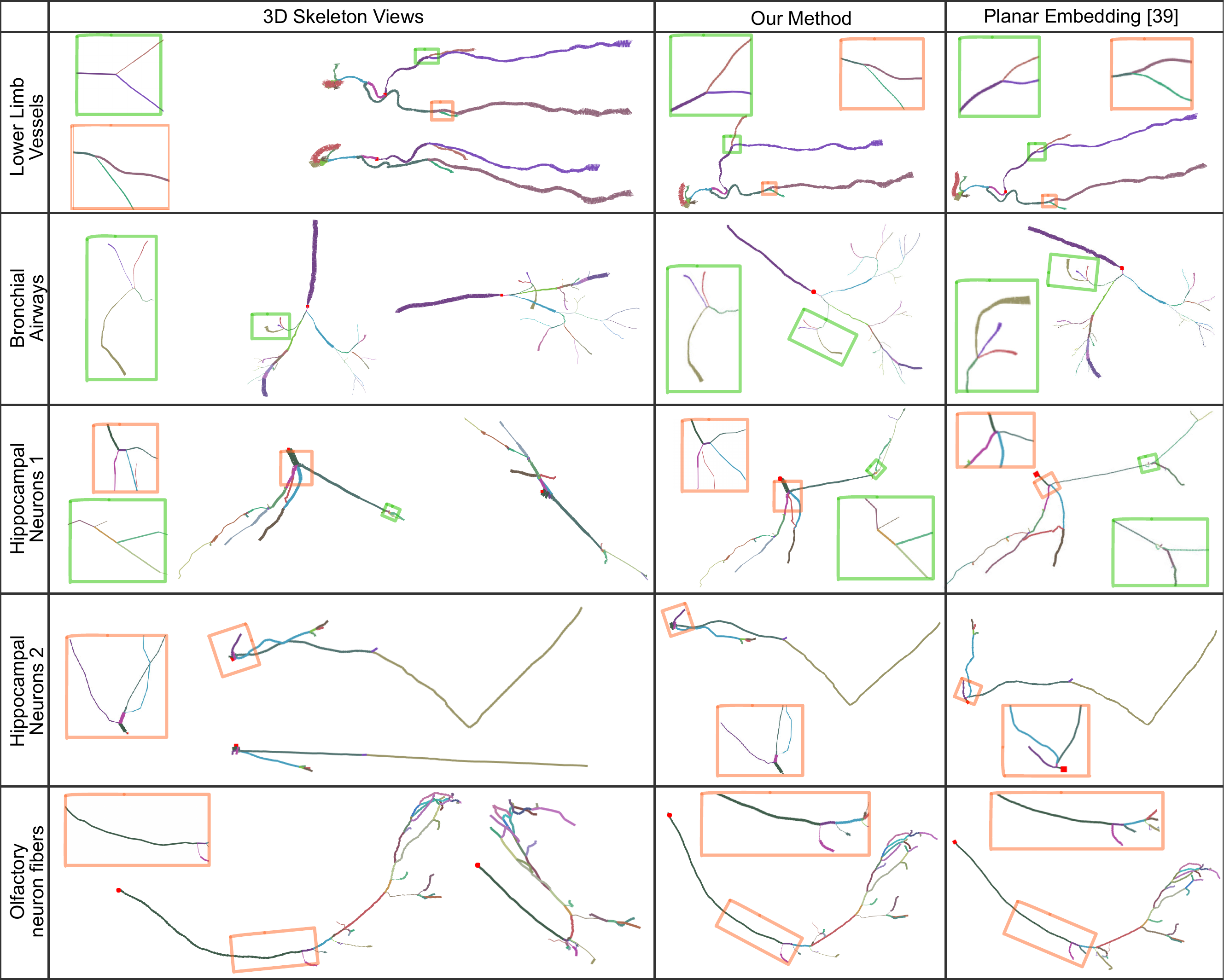}
\caption{
Planar embeddings using our method and the state-of-the-art method~\cite{marino2015planar}. 
For shape perception, the left column shows two 3D views of the skeleton with the top / left view being most representative chosen by a domain expert. 
We use colors to associate the respective segments in 3D and their planar representations. The thickness of the edges represents the node radii. 
The red square on each structure denotes the skeleton root node. 
For a closer comparison, the insets highlight the difference in geometry attributes preservation between our planar embedding result and the result using \cite{marino2015planar}, compared to the original 3D view.
Note that since \cite{marino2015planar} requires an input view, for a fair comparison, we generate the planar embedding from \cite{marino2015planar} using the expert-defined view.
}
\label{fig:results}
\end{figure*}

\begin{figure}
\centering
  \includegraphics[width=\linewidth]{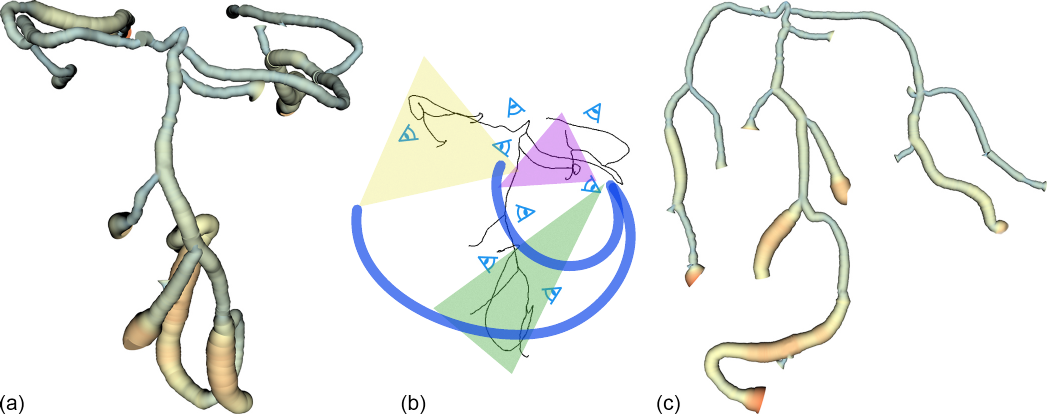}
 \caption{Our planar embedding method for the cranial blood vessels. (a) A 3D mesh rendering of an original structure of cranial blood vessels. (b) Illustration of three camera view fields focusing on a certain subtree structure. The blue curves represent a navigation example which connects the three hierarchical views. (c) Planar embedding result using our energy function that preserves the global shape of the object along with its local morphology, while avoiding intersections. The color represents the blood vessel radius of the segment with respect to the average vessel radius of the entire structure.}
	\label{fig:teaser}
\end{figure}

 \begin{figure}
    \centering
    \includegraphics[width = \linewidth]{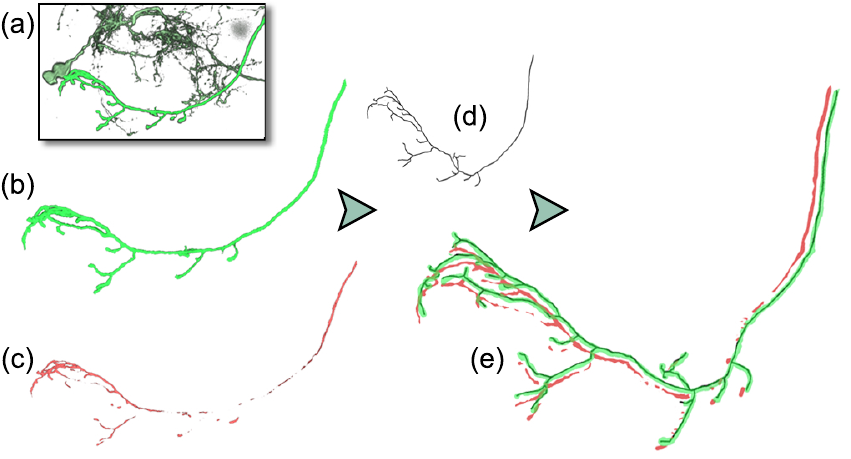}
    \caption{Juxtaposition of olfactory neurons for neurodegeneration pattern analysis using our planar embedding method.  (a) Brain microscopy volume.  (b) Neurite with its extensions which was extracted from the microscopy data.   (c) Corresponding predicted Alzheimer's disease structure~\cite{boorboor2021neuregenerate}.  (d) Planar embedding.  (e) Juxtaposition of neurite (b) and prediction (c) using the planar embedding (d) with an offset for clarity. }
    \label{fig:juxtapose}
\end{figure}


We demonstrate our work using examples from traced neuronal data (one olfactory and two hippocampal neurons from the DIADEM dataset~\cite{brown2011diadem}) and segmented vascular (cranial, lower limb, and aorta) and bronchial data.
The skeletons for the neuron datasets are extracted using gold-standard neuron tracing algorithms, provided with the datasets, and the skeletons for the blood vessels are computed using~\cite{bitter:2001:tvcg}.
Fig~\ref{fig:results} shows our planar embedding results for the five datasets, along with their 3D views and a comparison with embedding results generated using the state-of-the-art technique by Marino et al.~\cite{marino2015planar}. 
It can be seen here that a single 2D projection of a 3D structure introduces distortions and false intersections.
Whereas our embedding results can effectively preserve aspects of the original 3D shape using multiple local principal views.

\subsection{Planar Visualization}
The most simplistic visualization of an embedded structure is to render its projected skeleton as lines, similar to the rendered results in Fig.~\ref{fig:comparison}.  
Such a rendering provides a broad overview of the global and local morphology of a structure and can provide clear indications of specified locations along the skeleton.

A pseudo-surface rendering based on the radii values can be used to generate a 3D appearance of the embedded structure and also present a surface on which attributes can be mapped to color. Fig.~\ref{fig:teaser} shows such a surface rendering of segmented blood vessels from a CT scan. The color map\footnote{The diverging color map for three classes was designed using ColorBrewer~\cite{harrower2003colorbrewer} with the `color blind safe' and `print friendly' options.}, from a shade of blue to yellow to red, represents the gradient of the radii difference of a segment with respect to the average radius of the entire structure, in order of increasing segment radius value.  

A planar visualization for studying corresponding structural changes within a structure is possible by juxtaposing the projected structures in a single view.  We demonstrate this idea in Fig~\ref{fig:juxtapose} using an isolated \textit{healthy} olfactory neuron from a dense brain microscopy volume and its corresponding \textit{diseased} state predicted using~\cite{boorboor2021neuregenerate}. 
A planar embedding of the neuron's skeleton data (Fig.~\ref{fig:juxtapose}~(d)) is first constructed using our method.
Next, using a viewpoint perpendicular to the central axis of the skeleton, we render a bounded view volume using~\cite{boorboor2018visualization} and project it onto a single view along the embedding result. 
Finally, we juxtapose the two embedded volumes in a single view with an offset for clarity, as shownin Fig.~\ref{fig:juxtapose}~(e).

\subsection{Camera Navigation Path}
Exploring information in 3D space is the most natural way to fully investigate a 3D object.
With this motivation, our technique generates a camera path with dynamic focus on the different subtrees in the skeleton for users to automatically explore the data in 3D space. 
Fig~\ref{fig:exploration} demonstrates this concept for investigating a hippocampel neuron.
This camera path follows a top-down approach, starting with the best global view of the structure and then hierarchically navigating each subtree.
A spherical dolly motion around each subtree stimulates visual parallax to aid in understanding the geometry information along the view direction and a smooth transition between the two views gives the user a perception of the spatial relationship between different parts in the structure.
We refer the reader to our supplementary video for a more elaborate demonstration.


\begin{table}[hbt]
  \centering
  \caption{This table reports the computation time (in seconds) for our method and the method in~\cite{marino2015planar}. We ran our algorithm 5 times and report an average time for each dataset. The first four datasets are densely sampled skeletons while the remaining are relatively sparse. Our method is sensitive to the number of nodes and node density, thus leading to an increased computation cost with increasing node density.}
  \begin{tabular}{| c | c | c | c |}
  \hline
  {Dataset} & {Node number} & {Ours} & {\cite{marino2015planar}} \\
  \hline
  Lower Limb Vessels & 1538 & 212.5 & 11.5 \\
  \hline
  Cranial Blood Vessels & 1831 & 301.8 & 16.5 \\
  \hline
  Aorta Vessels & 2810 & 90.3 & 61.3 \\
  \hline
  Bronchial & 1031 & 119.9 & 4.5 \\
  \hline
  Hippocampal Neurons 1 & 175 & 11.6 & 17.8 \\
  \hline
  Hippocampal Neurons 2 & 101 & 5.1 & 12.1 \\
  \hline
  Olfactory Neurons & 235 & 19.2 & 13.9 \\
  \hline
  \end{tabular}
  \label{tab:timecost}
\end{table}

\section{Evaluation and Discussion}
\label{sec:eval}

\begin{table*}
  \centering
  \caption{Angle loss comparison between our method and the planar embedding method in \cite{marino2015planar}. }
  \begin{tabular}{| p{3.0cm}<{\centering} | p{0.55cm}<{\centering} |  p{0.55cm}<{\centering} |  p{0.55cm}<{\centering} |  p{0.55cm}<{\centering} | p{0.55cm}<{\centering} | p{0.55cm}<{\centering} | p{0.55cm}<{\centering} | p{0.55cm}<{\centering} |}
  \hline
 \multirow{3}*{Dataset}& \multicolumn{4}{|c|}{Metric 1: Projected Angle} & \multicolumn{4}{|c|}{Metric 2: Angle in 3D}\\
 \cline{2-9}
 ~ &\multicolumn{2}{|c|}{Max~$L(a)$} & \multicolumn{2}{|c|}{Avg~$L(a)$} &\multicolumn{2}{|c|}{Max~$L(a)$} & \multicolumn{2}{|c|}{Avg~$L(a)$}  \\
  \cline{2-9}
  ~ & Ours & \cite{marino2015planar} & Ours & \cite{marino2015planar}& Ours & \cite{marino2015planar}& Ours & \cite{marino2015planar} \\
  \hline
  Lower Limb Vessels & 0 & 1.560 & 0 & 0.788 & 0.140 & 5.818 & 0 & 0.016\\
  \hline
  Cranial Blood Vessels & 0.164 & 4.597 & 0.001 & 0.663 & 0.770 & 2.546 & 0 & 0.021 \\
  \hline
  Aorta Vessels & 0.090 & 5.279 & 0.001 & 0 & 0.634 & 2.951 & 0.005 & 0.082\\
  \hline
  Bronchial & 0.001 & 4.395 & 0 & 0.996 & 1.097 & 1.923 & 0 & 0.047 \\
  \hline
  Hippocampel Neurons 1 & 2.901 & 1.785 & 0.017 & 0.658 & 1.847 & 3.091 & 0.005 & 0.013 \\
  \hline
  Hippocampel Neurons 2 & 0.119 & 3.528 & 0.001 & 0 & 0.248 & 2.225 & 0 & 0.010 \\
  \hline
  Neuron Olfactory & 0.275 & 7.919 & 0.002 & 0 & 0.354 & 2.049 & 0 & 0.029 \\
  \hline
  \end{tabular}
  \label{tab:length_and_angle_loss}
\end{table*}

Our experiments were performed using a PC with an Intel Xeon Gold 6242 CPU @ 2.80GHz, 64 GB RAM, and one NVIDIA Quadro RTX 6000 graphics card.
The PSO algorithms were implemented in CUDA 11.2 such that each GPU thread is responsible for a solution candidate optimization. 
For our experiments, we set the following parameters: 40,960 swarm particles, $w_l = 2.0$, $w_a = 2.0$, $w_X = 0.2$, $\omega_g = 0.05$, $\omega_p = 0.05$, and $\omega_{inert} = 0.0375$. 
The performance of our technique is reported in Table~\ref{tab:timecost} and compared to the method in~\cite{marino2015planar}, tested in the same environment.

\subsection{Geometry Information Loss}
We provide a quantitative and qualitative evaluation of our planar embedding results and camera path generation.
Moreover, we compare our results with the skeleton embedding method in \cite{marino2015planar}.
\subsubsection{Quantitative analysis}
\label{sec:quantitative}
Table~\ref{tab:length_and_angle_loss} reports the geometry loss analysis using two metrics.

\textbf{Metric 1. Projected angle and length:} 
We evaluate the loss $L_l$ and $L_a$ by comparing our embedding with the target edge lengths and wth the target node angles, respectively:
$$L_l = \sum_{i}^{N-1}\frac{|l_i-l_{ei}|}{l_{i}} \quad \text{,} \quad L_a = \sum_{i}^{N}\frac{|\theta_{i} - \theta_{ei}|}{\theta_{i}} $$
where $N$ is the number of nodes.

\textbf{Metric 2. Angle and Length in 3D space:} 
Since the ultimate goal of planar embedding is to preserve the 3D geometry information, we also compute the edge length and angle loss compared to the original 3D structure.
In the original structure, each node angle is the acute angle between the edge and its parent edge.
The edge length and angle loss for the root node is 0.
We show that our technique can maintain the segment length well and the overall angle loss is small.

\begin{figure*}[t!]
\includegraphics[width=\textwidth,keepaspectratio]{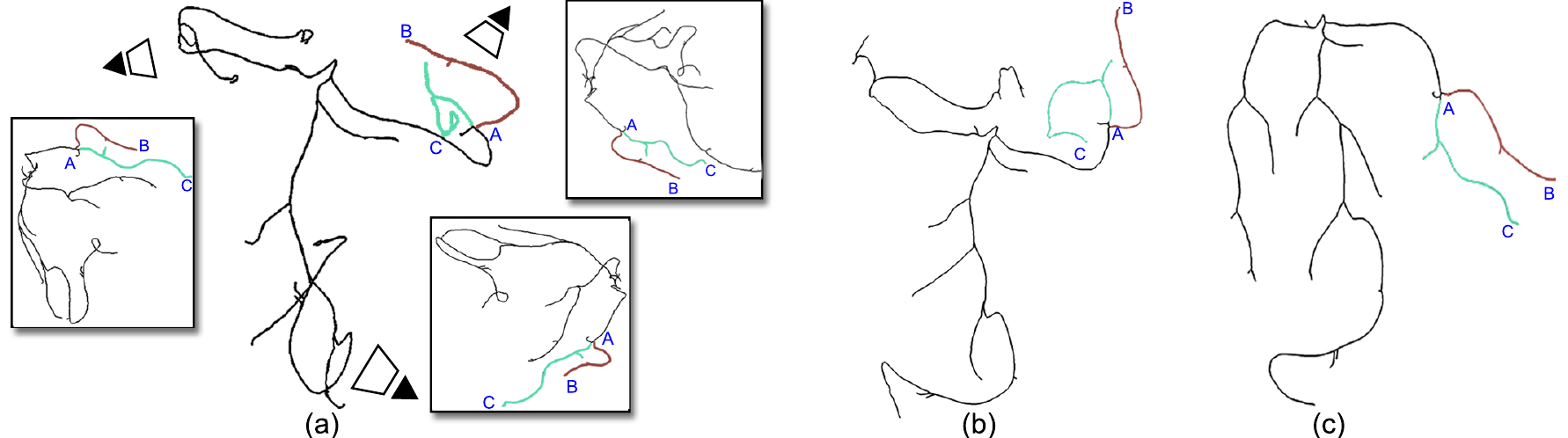}
  \caption{Preservation of local morphology in our work compared to a global-view based technique \cite{marino2015planar}.  (a) Different camera views of the structure, which may result in intersection and distortion of other branches, therefore making it difficult to find a global view that best captures the morphology of all branches.  (b) Embedding result from \cite{marino2015planar} that uses the global projection, resulting in misleading distortions highlighted using the teal-colored branch AC.  (c) Embedding using our method, which more accurately captures the local branching morphology.}
	\label{fig:comparison}
 \end{figure*}
 

\subsubsection{Qualitative Analysis}

As mentioned in Section~\ref{Sec:related_works_planar}, we consider~\cite{marino2015planar, won2012uncluttered} as methods closest to our work and ~\cite{marino2015planar} as the state-of-the-art.
The algorithm in these methods greatly depend on an appropriate input view, thus requiring users to manipulate 3D camera for an optimal input view. 
This is a challenging task that has motivated our work to require no user-defined input view.
Furthermore, the embedding from these method only demonstrate limited geometry information in the original 3D skeleton.
Fig.~\ref{fig:results} shows that our method outperforms ~\cite{marino2015planar} regarding visualizing segment shapes, especially when the input structure expands in all three dimensions in space, e.g., the bronchial airways, the cranial blood vessels, and the dense neuron fibers in olfactory neurites.

Fig.~\ref{fig:comparison} shows how a single global view can lead to misleading distortions.
Specifically for complex treelike objects where branching structures have varying principal axes, reducing the image space dimensionality can project the structure in a way that can distort its local morphology. 
The central structure in Fig.~\ref{fig:comparison}~(a) is a global projection of cranial blood vessels used in \cite{marino2015planar} to generate a planar embedding, and the corresponding insets show alternate projections from different 3D camera viewpoints.
This global view gives a perception that the branch AC has a \textit{curled} morphology. However, by observing the same branch from different viewpoints it can be seen that the AC has a much smaller curvature.
In comparison, since our embedding result uses local projection views to maintain local morphology and spatial relation with neighboring nodes, we see that the embedded segment AC in our result, shown in Figure~\ref{fig:comparison}~(c), does not have the misleading distortion.


\subsection{Expert Feedback}
We presented our technique to five experts (A1, A2, A3, A4, A5) from different research domains to comprehensively evaluate our method. 
Among the experts, A1 was a scientist in scientific visualization studying brain connectomics, A2 was a graduate student pursuing Doctor of Medicine who worked on medical visual analytics, A3 was a neuroscientist focusing on microscopy biology data including neurons, A4 was a doctor in nutritional science who heavily uses hierarchical visualization tools to organize datasets and conducts data mining, and A5 was a scientist in single cell genomics, studying the evolutionary trajectories of plant cells.

Each expert participated in this evaluation separately. 
They were presented with the planar embedding of interested datasets, the interactive embedding GUI, and the automatic exploration in 3D space, followed by an interview to collect their feedback.
A1 observed the neocortical axon structure and the bronchial airways; A2 observed the lower limb vessels, the upper aorta vessels, the olfactory neuron fibers, and the neocortical axon structure; A3 examined the neocortical axon structure; A4 observed the bronchial airways and the olfactory neuron fibers; and A5 studied one hippocampal neuron model and the olfactory neuron fibers.

The interview started with a description of treelike structure visualization.
Next, the experts were asked three profile questions: ``Do you have any experience of studying treelike structures datasets?"; ``When studying your data, what are your focused attributes?"; and ``Can you give a score (on a scale of 1 as no expertise, to 5 as authoritative understanding and experience) to rate your expertise on general data visualization and data visualization with treelike structures, respectively?".
Experts' self ratings for the first and third questions are (5, 4, 5, 3, 4) and (4, 3, 5, 2, 2), respectively.
For 3D navigation, experts chose the hierarchical views and navigated themselves using a GUI slider to control the speed.
We then asked ``What features are helpful for your research?" and ``What can be improved to better assist your research?".
Additionally, participants was asked to rate the overall practicality of our method and the two major functions, i.e., planar embedding with interactive edit, and suggested 3D exploration path. 
A1, A2, A3, and A4 rated the overall method and the embedding as 4. A5 rated the overall method as 5.
For the itemized score, A1 rated the navigation as 5 and the others rated it as 4.

The feedback from the experts were positive.
All experts appreciated the skeleton embedding with segment shape preservation.
According to A2, in vascular studies, it is important to reveal the vessel's branching pattern and connections since they are unique for each person.
A3 noted that ``This technique can be very useful in studying neuron sample's structure".
A5 highly rated the planar embedding because ``the method can be very useful for studying plant root system".
A1, A2, A4, and A5 favored the camera dolly motion around each hierarchical view in the path.
A1 said, ``This is helpful for understanding the neural connections coming from different areas in the brain''.
A2 preferred our LoD function for 3D treelike structure.
Both A1 and A3 emphasized that they would like to use the 3D exploration and 2D embedding visualization together because the combined modes can invoke different insights.
A5 commented that this effective visualization tool simplifies the study of complex root structures.
Regarding what can be improved, A1 commented that the sub-structures in 3D navigation view should be dynamically highlighted in the 2D embedding to enhance structural understanding.
A2 suggested to tune the distance between the camera and the center of the data and to adjust the visualized width of the segment in the skeleton structure for aesthetic purposes.
A4 suggested indicating the existing manual edits on the embedding to help researchers be aware of the accumulated changes.
\subsection{Discussion and Limitations}
\begin{figure}
\centering
\includegraphics[width=\linewidth]{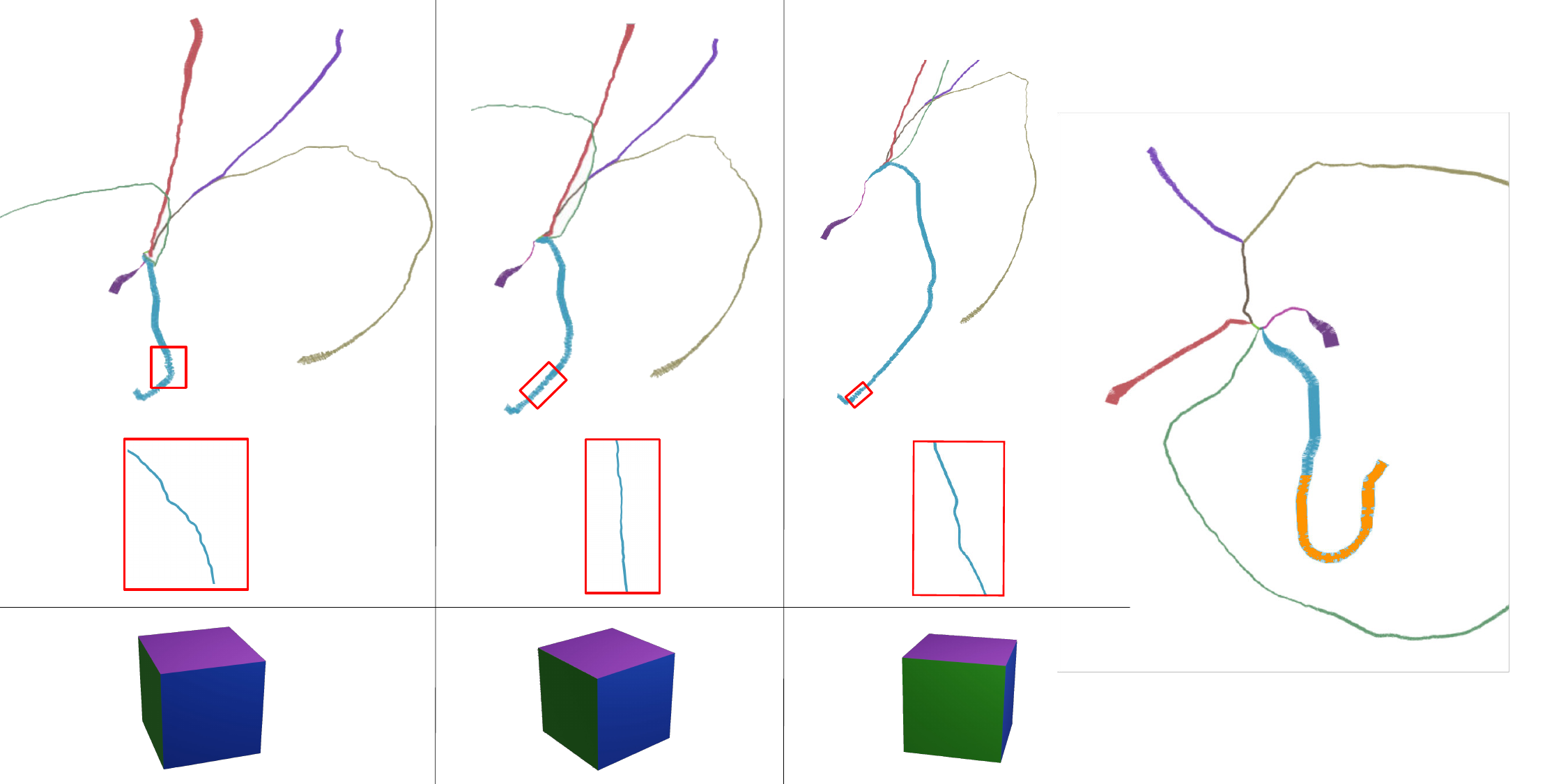}
\caption{Local shape preservation and an example of avoiding intersections. Left: details of the node angles are shown by zooming in. The cube below each subfigure shows the relative transformation between each view. Right: in the blue curve, the bending near the end is caused by the potential conflict with the green curve below. Our method adjusts the expansion scope of the blue curve based on the target of optimizing the sum of the geometry loss. }
\label{fig:aorta_explain}
\end{figure}

Our planar embedding method preserves segment length, angles, and high level shape, especially revealing detailed node angles that are easily overlooked by people in global view observation.
Fig.~\ref{fig:aorta_explain} demonstrates an example of this. Although the blue curve in the upper aorta vessel structure appears to be quite straight in a global view, we can observe bending when zoomed into the subtree.
However, since our method avoids intersection, some large bending angles may be introduced during optimization, when a curve's expansion conflicts with other curves.
For example, we notice a bending near the of the blue curve in Fig.~\ref{fig:results}.
This is caused by the conflict between the expansion scope of the blue curve during the green curve's placement below it.

While we optimize an exhaustive list, more attributes that preserve treelike structures need to be explored.
For instance, relative position between different segment pairs may be a useful geometry attribute in some research domains. Since each segment is projected onto the principal plane which is determined by the hierarchical view, the relative location of segments on a projection of an arbitrary view is not always consistent with the view of the principal plane.

\subsubsection{Parameter analysis}
\begin{figure}
\centering
\includegraphics[width=\linewidth]{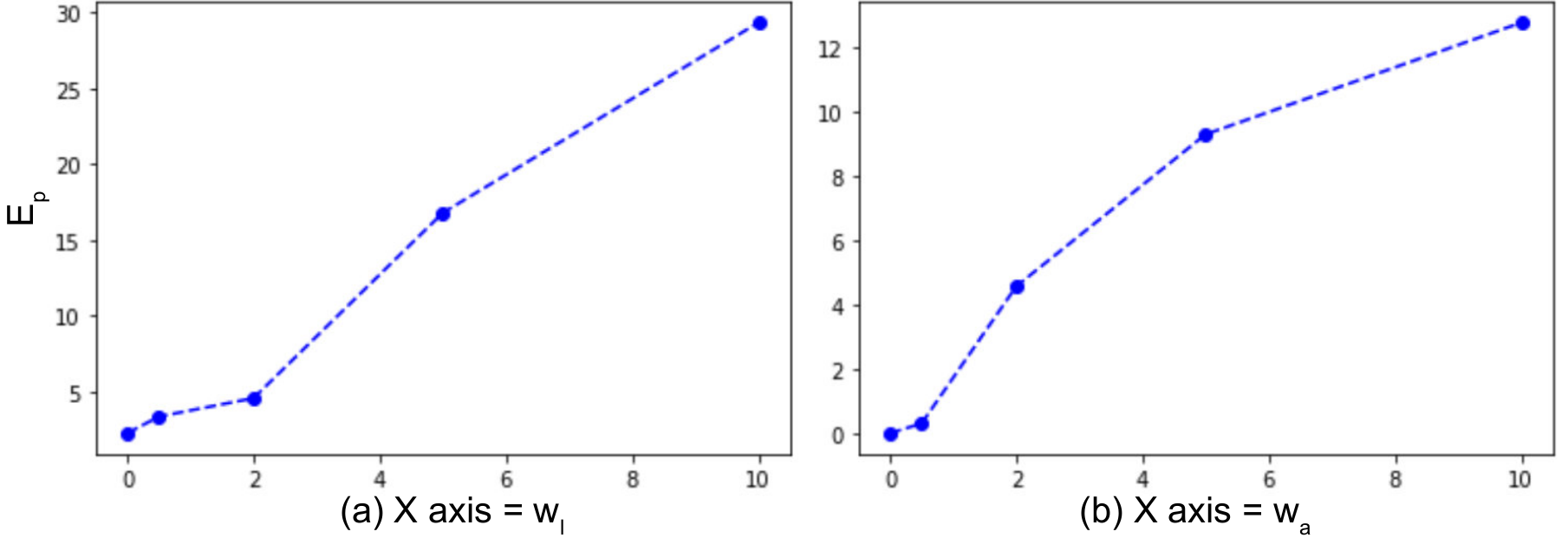}
\caption{The relationship between the embedding's geometry loss and the parameters $w_l$, $w_a$ in the optimization. Left: $w_a=2.0$. Right:$w_l=2.0$}
\label{fig:rarl}
\end{figure}

The combination of $w_l$ and $w_a$ reflects the importance of length and angle preservation.
With a fixed $w_a$, increasing $w_l$ yields a smaller edge length loss.
Similarly, fixed $w_l$ and increasing $w_a$ results in a smaller angle loss.
Fig.~\ref{fig:rarl} demonstrates geometry losses for $w_l$ and $w_a$ combinations, using Metric 1 on Hippocampal Neurons 1.

The parameter adjustment allows the users to assign varying $w_l$ and $w_a$ for different nodes based on their specific visualization needs.
For instance, to impose strict length preservation closer to the root node, the nodes with the shorter depth can be assigned a higher $w_l$ than other nodes in the tree.

\subsubsection{Limitations}
The PSO time complexity is $O{(N^2)}$. The performance bottleneck is the segment intersection checking task in each loop. This intersection checking is composed of a coarse and fine intersection checking. The coarse check facilitates faster computation when the nodes are sparsely placed. As shown in Table~\ref{tab:timecost}, when optimizing the planar embedding for structures with dense nodes, our time cost is more expensive than the method in~\cite{marino2015planar}. 

One failure case is caused by extremely large radii of the tree segments. Since our embedding preserves the edge radius, the intersection checking stage filter out the conditions when wide edges overlap with each other. Therefore, it is guaranteed that when applying the surface mesh onto the skeleton, no occlusion will be introduced. 
In the cases with extremely large radius there may not even be a valid radial layout embedding, thus resulting in no solutions for the planar embedding.

\subsubsection{Applications}
\label{sec:application}

It has been shown that 2D representation of medical data helps carry out tasks more effectively~\cite{borkin2011evaluation} and therefore, by preserving the shape of the original structure, a domain expert can more efficiently navigate the 3D structure to confirm the findings from the planar view.
Preserving the shape also allows domain scientists to present the representation to collaborators who will be able to recognize the anatomy.
In this paper, we used neurite analysis and blood vessel visualization as two examples to demonstrate the utility of our method.
\par
\textbf{Vascular analysis. }Though 3D rendering already provides a better view than na\"ively scrolling through slices of the CT scan, its visualization efficiency depends on the user's ability to seamlessly manipulate camera views within the application to study the entire structure and simultaneously make a mental note of features such as radii and abnormalities.
In contrast, using a planar visualization as in Fig.~\ref{fig:teaser}(c), the user is able to visualize all of the blood vessels and compare their geometric attribute (i.e., radius) within a single view. 
Moreover, at an application level, we have shown that our camera view technique can be used to generate viewpoints that results in minimal visual occlusion and maximizes the spatial understanding of the structure. 

\par
\textbf{Neurite analysis. }
Connectomics~\cite{TheHumanConnectome} is an emerging field for techniques that study complex neural connection maps.
By mapping the brain connectivity, neuroscientists will be able to analyze how the human brain functions and its degradation process as a result of cognitive decline or disease~\cite{ghahremani2021neuroconstruct}.
A common challenge in this domain is the immensity of volume data. 
Recent works~\cite{mohammed2017abstractocyte,al2014neurolines} have introduced methods to alternatively represent dense neuronal structures as abstracted diagrams, albeit in a linear graph-like arrangement.
Though schematic diagrams help study connections and topological layouts, an examination of the shape of the structure is required to understand morphological changes during degeneration. 
Moreover, given the density of information within a microscopy volume and the intricate neurite morphology, neuroscientists independently observe regions of the brain and make general visual observations using population analysis.  
Our embedding and visualization method can allow for effective analysis of fragmentation and thinning patterns within a neuron by juxtaposing the \textit{healthy} and \textit{diseased} neurites, as shown in Fig~\ref{fig:juxtapose}~(e), instead of separately visualizing them in 3D.

\section{Conclusion and Future Work}
\label{sec:conclusion}

We have presented a new method of generating geometry-aware planar embeddings of treelike structures based on their skeletons.  Unlike previous methods, our PSO implementation does not require an input view position, and it also accounts for the local morphology of the object.  
The user can interactively adjust branch locations as desired, and the embedding can be updated based on these modified positions. 
The local camera projection views generated to create the embedding can also be used for exploration of the original structure in 3D, providing optimal views of the various subtrees. 
We have demonstrated and evaluated our technique using a variety of real-world data, including blood vessels, bronchi, and traced neuron data.
Currently, our work is limited to embedding the skeleton of the structure in 2D.  While a pseudo-surface rendering based on the radius values for each node can give a sense of a solid structure, we plan to explore in the future how the actual surface or volume data can be incorporated into the embedding.  We will also work with domain experts to further refine our results and incorporate our method into an intuitive user interface which addresses the needs specific to different fields.

\section*{Acknowledgements}
This project was supported in part by NSF grants CNS1650499, OAC1919752, ICER1940302, and IIS2107224.
Bronchial CT scans courtesy of the EXACT’09 study, lower limb vessels courtesy of OsiriX, upper aorta courtesy of the Visible Korean Human Project, and cranial vessels courtesy of VolVis.

\ifCLASSOPTIONcaptionsoff
  \newpage
\fi


\bibliographystyle{abbrv-doi}

\bibliography{main.bib}

\begin{IEEEbiography}[{\includegraphics[height=1.25in,width=1in,keepaspectratio]{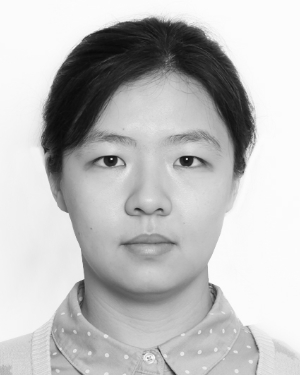}}]{Ping Hu}
is currently pursuing the PhD degree in Computer Science at Stony Brook University. 
She received her Bachelor degree of Physics at Shandong University, China.
Her research focuses on computational camera control in scientific visualization and computer graphics. 
\end{IEEEbiography}

\begin{IEEEbiography}[{\includegraphics[height=1.25in,width=1in,keepaspectratio]{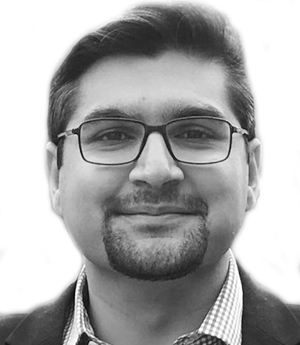}}]{Saeed Boorboor}
is currently pursuing a PhD degree in  Computer Science at Stony Brook University. He received his BSc Honors in Computer Science from School
of Science and Engineering, Lahore University of Management Sciences, Pakistan. His research interests include scientific visualization, biomedical imaging, and computer graphics.
\end{IEEEbiography}


\begin{IEEEbiography}[{\includegraphics[height=1.25in,width=1in,keepaspectratio]{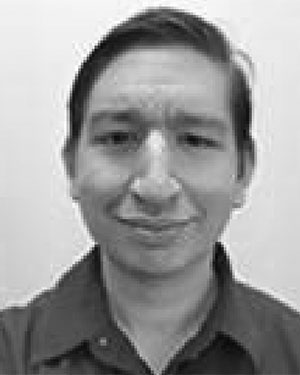}}]{Joseph Marino}
is a Postdoctoral Research Fellow.
He received his PhD in Computer Science
(2012), and BS in Computer Science and Applied
Mathematics and Statistics (2006) both from
Stony Brook University. His research focuses on
visualization and computer graphics for medical
imaging, including visualization and analysis for
pancreatic and prostate cancer screening, and
enhancements for virtual colonoscopy. He is a
recipient of the 2016 Long Island Technology
Hall of Fame Patent Award (2016).
\end{IEEEbiography}

\begin{IEEEbiography}[{\includegraphics[keepaspectratio]{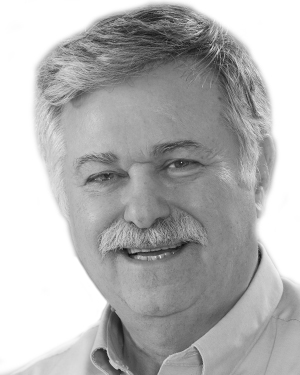}}]{Arie E. Kaufman} is a Distinguished Professor of Computer Science, Director of Center of Visual Computing, and Chief Scientist of Center of Excellence in Wireless and Information Technology at Stony Brook University. He served as Chair of Computer Science Department, 1999-2017. He has conducted research for $>$40 years in visualization and graphics and their applications, and published $>$350 refereed papers. He was the founding Editor-in-Chief of IEEE TVCG, 1995-98. He is an IEEE Fellow, ACM Fellow, National Academy of Inventors Fellow, recipient of IEEE Visualization Career Award (2005), and inducted into IEEE Visualization Academy (2019) and Long Island Technology Hall of Fame (2013). He received his PhD in Computer Science from Ben-Gurion University, Israel (1977).
\end{IEEEbiography}

\end{document}